\documentclass[11pt]{article}
\usepackage[fleqn]{amstex}
\usepackage{amsthm,amssymb}
\usepackage{cite}

\newtheorem{prop}{\sc Proposition}

\newtheorem{lemme}{\sc Lemma}

{\theoremstyle{remark}
\newtheorem{rem}{Remark}}

\def\cqfd{\hfill\nobreak\hbox{${}_\Box$}\par\medbreak}

\newcommand{\esp}{\qquad\!}

\newcommand{\ket}[1]{|\,#1\,\rangle}

\def\Id{\mathrm{Id}}

\def\tr{\operatorname{tr}}

\def\End{\operatorname{End}}

\def\sn{\operatorname{sn}}
\def\cn{\operatorname{cn}}
\def\dn{\operatorname{dn}}

\def\le{\leqslant}

\def\pour#1{_{\,\vrule height 13pt depth 1pt\> {#1}\!}}

\def\build#1_#2^#3{\mathrel{
\mathop{\kern 0pt#1}\limits_{#2}^{#3}}}

\newcommand{\Cset }{{\mathbb C}}

\def\cH{\mathcal{H}} 
\def\cF{\mathcal{F}}

\def\cL{\mathcal{L}}

\usepackage{euscript}

\renewcommand{\H}{{\boldsymbol{\mathrm{H}}}}

\def\de{\delta}

\def\eps{\varepsilon}

\def\la{\lambda}
\def\La{\Lambda}

\begin{document}


\begin{titlepage}
\begin{flushright}
LPENSL-TH-19/99\\
\end{flushright}
\par \vskip .1in \noindent

\begin{center}
{\LARGE On the quantum inverse scattering problem}\\
\end{center}
  \par \vskip .3in \noindent

\begin{center}

      {\bf J. M. MAILLET and V. TERRAS}
  \par \vskip .1in \noindent

  Laboratoire de Physique $^{**}$\\
Groupe de Physique Th\'eorique\\
       ENS Lyon, 46 all\'ee d'Italie 69364 Lyon CEDEX 07
       France\\[0.6in]
\end{center}

\par \vskip .10in
\begin{center}
{\bf Abstract}\\
\end{center}

\begin{quote}
A general method for solving the so-called quantum inverse scattering problem (namely the reconstruction of local quantum (field) operators in term of the quantum monodromy matrix satisfying a Yang-Baxter quadratic algebra governed by an $R$-matrix) for a large class of lattice quantum integrable models is given. The principal requirement being the initial condition ($R(0) = P$, the permutation operator) for the quantum $R$-matrix solving the Yang-Baxter equation, it applies not only to most known integrable fundamental lattice models (such as Heisenberg spin chains) but also to lattice models with arbitrary number of impurities and to the so-called fused lattice models (including integrable higher spin generalizations of Heisenberg chains). Our method is then applied to several important examples like the $sl_n$ $XXZ$ model, the $XYZ$ spin-$1 \over 2$ chain and also to the spin-$s$ Heisenberg chains. 

\end{quote}

\begin{flushleft}
\rule{5.1 in}{.007 in}\\
$^{**}${\small UMR 5672 du CNRS, associ\'ee \`a  l'Ecole
Normale Sup\'erieure de Lyon.}\\
{\small This work is supported by CNRS (France), the EC-TMR contract FMRX-CT96-0012, MAE fellowship 96/9804 and MENRT (France) fellowship 
AC 97-2-00119.}\\
{\small email: Jean-Michel.Maillet\symbol{'100}ens-lyon.fr, Veronique.Terras\symbol{'100}ens-lyon.fr}\\[0.2 in]

\end{flushleft}

\end{titlepage}



\section{Introduction}

One of the main challenging problem in the theory of quantum integrable models, after diagonalizing the corresponding Hamiltonians, is to find explicit and manageable expressions for their correlation functions. This is well known to be a very hard problem in general situations, even for quantum integrable models in two dimensions, and in fact besides the models which can be connected to free fermions or which are in the conformal field theory class, very few explicit solutions to this problem have been obtained so far.

For example, in the case  of the Heisenberg spin-$1 \over 2$ chains \cite{Hei28}, while the method for diagonalizing the Hamiltonian was designed in 1931 by H. Bethe \cite{Bet31} (for the isotropic XXX case), the first manageable expressions for its correlation functions (in the thermodynamic limit) have been given only in 1992 (in the massive regime) \cite{JimM95L} and then in 1996 (in the gapless regime) \cite{JimM96} by Jimbo, Miwa and their collaborators, however based on some nice hypothesis, in particular, on the symmetry structure of the infinite chain.

Motivated by this problem, it was shown for the first time in \cite{KitMT99} that the explicit solution of the quantum inverse scattering problem (namely the reconstruction of any local spin operator at any site of the chain in terms of the elements of the quantum monodromy matrix satisfying a Yang-Baxter algebra, and containing creation/annihilation operators of Bethe eigenstates of the Hamiltonian) can be solved for this $XXZ$ spin-$1 \over 2$ model in a very simple (multiplicative) way.  Moreover, using the algebraic Bethe ansatz framework \cite{FadST79,FadT79}, it was then used in \cite{KitMT99a} to derive explicit and simple expressions for the correlation functions of this model, even in the presence of a non-zero magnetic field.  In the thermodynamic limit (the limit of infinite chain) it leads  to multiple integral representations of these correlation functions. In the zero magnetic field limit, this approach gives then a complete proof of results and conjectures previously obtained in \cite{JimM95L,JimM96}.

The elementary nature of the answer to the quantum inverse scattering problem, in this very representative example among the models solvable by means of the algebraic Bethe ansatz method, was quite unexpected. The algebraic Bethe ansatz method, also called quantum inverse scattering method, appeared  20 years ago as a quantum analogue of the classical inverse scattering problem approach to non-linear wave equations having soliton solutions, in order to solve quantum integrable models in two dimensions. The essential tools of this method are the quantum monodromy matrix satisfying quadratic commutation relations (Yang-Baxter algebra) which structures constants are given by an $R$-matrix solving the Yang-Baxter (cubic) equation \cite{Bax82L,Yan67}. It has been designed to diagonalize the corresponding Hamiltonians  simultaneously with their associated commuting family of integrals of motion \cite{FadST79,FadT79}. It is the quantum analogue of the direct part of the classical inverse scattering problem method in its Hamiltonian formulation \cite{ZakF71}, in which the Lax matrix is used to construct the monodromy matrix containing the action-angle variables which linearize the dynamics. However, the inverse scattering problem part of the classical theory, namely, the reconstruction of the local classical field variables contained in the Lax matrix in terms of the elements of the monodromy matrix (and hence in terms of the action-angle variables) using the Gel'fand-Levitan-Marchenko equations \cite{GelL55,Mar55,AblKNS73}, being already a quite difficult problem to solve, it was not at all obvious to find a direct way to extend it to the quantum situation, although the motivations (form factors and correlation functions) were very clear from the very beginning \cite{CreTW80,CreTW80a,CreTW81a}.

The very remarkable feature of the solution to this problem given in \cite{KitMT99} was not only that the quantum inverse scattering problem can be solved explicitely, but also that both its resulting expressions (reconstruction of the local spin operators at any site of the chain in terms of a simple product of the quantum monodromy matrix elements) and their proofs are very elementary. In turn, it essentially relies on the fact that the quantum $R$-matrix $R(\lambda, \mu)$ solving the Yang-Baxter equation  reduces to the permutation operator when the two spectral parameters $\lambda$ and $\mu$ are equal. 

This fact being almost a consequence of the Yang-Baxter equation itself, and hence satisfied for very generic cases, it immediatly suggested that the quantum inverse scattering problem can indeed be solved for almost all known quantum integrable (lattice) models\footnote{This idea was given by the authors in various conferences, including the``Seminaire de Math\'ematique sup\'erieure" held in Montreal in the summer 1999 \cite{Mai99}.}. This is to be compared to the tentatives of solving the corresponding problem directly for the continuum quantum integrable field theories such as the Sine-Gordon model, using a quantum version of the Gel'fand-Levitan-Marchenko equations, where it appeared to be an extremely difficult problem to handle, due in particular to the presence of boundstates \cite{Smi86,Smi92L}. 

The results we will present here show that for lattice models at least (whatever complicated is the corresponding spectrum) this problem admits a simple and purely algebraic solution.

More precisely, let us consider a quantum lattice model associated to the finite dimensional representations of some quantum algebra $\mathcal{A}$ \cite{Jim85,Dri85}, and let $W_r(\la)$, where $r+1$ is the dimension, such representations depending on a spectral parameter $\la$ (here we have in mind for example the finite dimensional evaluation representations of some quantum affine algebra). Let us suppose that the corresponding quantum $R$-matrix $R (\la, \mu)$ is invertible in any irreducible tensor product $\End(W_{r_n}(\la)\otimes W_{r_n}(\mu))$ and reduces to the permutation operator for equal spectral parameters.
Then, let the quantum Lax operator $L_n (\la)$ at site $n$ defining this lattice model be obtained as some finite dimensional representations of the quantum  $R$-matrix in $\End(V(\la)\otimes W_{r_n}(\xi_n))$ as,
\begin{align*}
    L_n(\la-\xi_n+\nu_0) &\equiv L^{VW_{r_n}}_{0n}(\la-\xi_n+\nu_0)\\
                  &= R^{V W_{r_n}}_{0n}(\la-\xi_n)
      \ \in \End(V(\la)\otimes W_{r_n}(\xi_n)),
\end{align*}
where the auxiliary space $V(\la)$ (labelled by subscript 0) is for example  a fondamental (evaluation) representation of $\mathcal{A}$, $W_{r_n}(\xi_n)$ is the quantum local space at site $n$, and $\xi_n$ is an inhomogeneity parameter attached to the site $n$, the space of states of the model being given by,
\begin{equation*}
     \cH \simeq W_{r_1}(\xi_1)\otimes\dots\otimes W_{r_N}(\xi_n).
\end{equation*}
Let $L_n (\la)$ be choosen moreover such that we can associate to it an auxiliary quantum Lax operator ${\cal{L}}_n (\la)$ obtained by fusion \cite{KulRS81,Jim85,Jim86,KirR87} of $L_n (\la)$ (meaning that its operator matrix elements are functions of the one's contained in $L_n (\la)$) and given by the quantum $R$-matrix in $\End(W_{r_n}(\la)\otimes W_{r_n}(\xi_n))$. Then, even in the general situation where the local quantum space of states $W_{r_n}(\xi_n)$ are different from one site to another,  it is possible to give an explicit and elementary solution of the reconstruction of the local operators at any site $n$ of the lattice in terms of elements of the quantum monodromy matrix associated to $L_n (\la)$. In fact the resulting expressions for these local operators reconstructions and their proofs are very direct generalizations of the one's given for the  Heisenberg XXZ case in  \cite{KitMT99} which is quite generic in this respect. 

>From a more universal point of view, which will be useful when considering the extension of our results to the field theory case for which infinite dimensional representations should be considered, the solution of the quantum inverse scattering problem relies in turn to the very simple fact that the associated universal $\cal{R}$-matrix is the canonical (identity) element in the corresponding double Hopf algebra $\cal{D} (\cal{A})$ \cite{Dri87}. 

The main purpose of this article is to give a simple and general solution to the quantum inverse scattering problem for any quantum integrable lattice model of the above type solvable by means of the algebraic Bethe ansatz method \cite{FadST79}. This result applies to most known quantum lattice integrable models, even with arbitrary number of impurities at different lattice sites, since our method takes into account the possibility to have  any chosen (finite dimensional) representations for the local space of states at each site which could be different for different sites. We will also give explicit formulas for  several known interesting examples including not only the so-called fundamental lattice models  (such as XYZ spin-$1 \over 2$ or $sl_n$ XXZ models, for which the result is essentially a direct copy of the case described in \cite{KitMT99}), but also for the more general cases of fused lattice models (such as the higher spin generalisations of the Heisenberg chains). In that last interesting situation, we will also discuss the implications of these results for the classical inverse scattering problem. 

The application of these results to the actual computation of correlation functions of the above models will be considered in forthcoming publications.

This article is organized as follows. In section 2, we first recall the results for the spin-$1 \over 2$ XXZ Heisenberg spin chain as given in \cite{KitMT99}, which will be a guideline all along the article. Then the results are extended in a quite direct way to a large class of lattice quantum integrable models in section 3. In section 4, we apply these results to the so-called fundamental lattice models for which the quantum Lax operator is given essentially by the corresponding quantum $R$-matrix, namely, for which in particular the auxiliary and quantum spaces are isomorphic. In section 5, we consider the so-called fused lattice models, with the example of the XXX spin-$s$ model. Some conclusions and comments are given in the last section.



\section{Solution of the quantum inverse problem for the Heisenberg spin-1/2 
chain}

In this section, we recall the solution of the quantum inverse problem which
was given in \cite{KitMT99} in the case of the XXZ spin-1/2 Heisenberg chain.

The XXZ Heisenberg chain of length $N$ corresponds to the following 
Hamiltonian,
\begin{equation}
    H_{\mathrm{XXZ}}=\sum_{m=1}^N \big\{ \sigma_m^x \sigma^x_{m+1} +
  \sigma^y_m\sigma^y_{m+1} + \Delta(\sigma^z_m\sigma^z_{m+1}-1)\big\},
\label{eq:H} 
\end{equation}
where $\sigma^\alpha_m,\ \alpha=x,y,z,$ are the Pauli matrices acting in the two-dimensional space $\mathcal{H}_m$ at 
site $m$, $\Delta=\cosh \eta$ is the anisotropy parameter,
the particular case $\Delta=1$ corresponding to the XXX chain. 
We impose periodic boundary conditions.

In the algebraic Bethe ansatz framework, the corresponding quantum $R$-matrix
is 
\begin{equation}\label{mat-R}
   R(\lambda, \mu)=
      \begin{pmatrix}
         1 & 0 & 0 & 0\\
         0 & b(\lambda, \mu) & c(\lambda, \mu) & 0\\
         0 & c(\lambda, \mu) & b(\lambda, \mu) & 0\\
         0 & 0 & 0 & 1
      \end{pmatrix}
\end{equation}
where
\begin{align}\label{matbc}
  b(\lambda, \mu) & = \frac{\varphi(\lambda-\mu)}{\varphi(\lambda-\mu+\eta)},\\
  c(\lambda, \mu) & = \frac{\varphi(\eta)}{\varphi(\lambda-\mu+\eta)},
\end{align}
with the function $\varphi$ defined as
\begin{alignat}{3}\label{def-phi}
   \varphi(\lambda) & = \lambda         &  &\text{ in the XXX }&\text{case,}\\
   \varphi(\lambda) & = \sinh(\lambda)\ &  &\text{ in the XXZ }&\text{case.}   
\end{alignat}
The $R$-matrix is a linear operator in the tensor product of two 
two-dimensional linear spaces $V_1 \otimes V_2$, where each $V_i$ is 
isomorphic to ${\Cset}^2$, and depends generically on (the difference of)
two spectral parameters $\lambda_1$ and $\lambda_2$ associated to these two 
vector spaces. 
It is denoted by $R_{12} (\lambda_1, \lambda_2)$ or by
$R_{12} (\lambda_1- \lambda_2)$. 
Such an  $R$-matrix satisfies the Yang-Baxter equation,
\begin{equation}
  R_{12} (\lambda_1, \lambda_2)\ R_{13} (\lambda_1,\lambda_3)\ 
      R_{23} (\lambda_2,\lambda_3) = 
  R_{23} (\lambda_2,\lambda_3)\ R_{13} (\lambda_1,\lambda_3)\ 
      R_{12} (\lambda_1,\lambda_2).
\end{equation}
Moreover, it is unitary, except for a finite number of  values of the
spectral parameter 
(such that $b(\lambda_1, \lambda_2) = \pm c(\lambda_1, \lambda_2)$),
\begin{equation}
  R_{12} (\lambda_1-\lambda_2)\ R_{21} (\lambda_2-\lambda_1) = {\mathbf 1},
\label{eq:eu}
\end{equation}
and reduces to the permutation operator for a particular value of the spectral parameter, $R_{12} (0) = P_{12}$

Identifying one of the two linear spaces in the $R$-matrix 
with the quantum local space ${\cal H}_n$ at site $n$, 
one obtains the quantum $L$-operator of the model at site $n$ as
\begin{equation}
 L_n(\lambda, \xi_n)=R_{0n}(\lambda- \xi_n),
\end{equation}
where $\xi_n$ is an arbitrary (inhomogeneity) parameter dependent on the site 
$n$.  The Hamiltonian \eqref{eq:H} corresponds to the homogeneous case
where all the inhomogeneity parameters $\xi_n$ are equals.
The subscripts mean here that $R_{0n}$ acts on the tensor product of the
auxiliary space $V_0\simeq\Cset^2$ by the quantum space ${\mathcal H}_n$
at site $n$.
The quantum monodromy matrix of the total chain is then defined as the 
ordered product of all $L$-operators along the chain,
\begin{equation}
T_0 (\lambda)\ \equiv\ T_{0, 1 \ldots N} (\lambda; \xi_1, \ldots ,\xi_N) =
      R_{0N}(\lambda- \xi_N)\ldots R_{01}(\lambda- \xi_1),
\end{equation}
and can be represented in the auxiliary space $V_0$ as a $2 \times 2$ matrix,
\begin{equation}\label{mat-monodromie}
   T (\lambda) = 
      \begin{pmatrix}
         A (\lambda) & B (\lambda)\\
         C (\lambda)& D (\lambda)
      \end{pmatrix}, 
\end{equation}
whose matrix elements $A$, $B$, $C$, $D$
are linear operators on the quantum space of states of the chain 
${\cal H}=\mathop\otimes\limits_{n=1}^{N}{\cal H}_n$.
Their commutation relations are
given by the following relation on ${\Cset}^2 \otimes {\Cset}^2$:
\begin{equation} \label{commutation}
   R_{12}(\lambda, \mu)\ T_1(\lambda)\ T_2(\mu)  
      = T_2(\mu)\ T_1(\lambda)\ R_{12}(\lambda, \mu),
\end{equation}
with the usual tensor notations $T_1(\lambda)=T(\lambda) \otimes \Id$ and
$T_2(\mu)=\Id \otimes T(\mu)$.

\bigskip

In the algebraic Bethe ansatz framework, the matrix elements of the monodromy
matrix are used to construct the eigenstates of the Hamiltonian.
In particular, the transfer matrices $t(\la)$, which are the trace
$A(\la)+D(\la)$ of the corresponding monodromy matrices \eqref{mat-monodromie},
commute with each other for different values of the spectral parameter $\la$,
and the Hamiltonian \eqref{eq:H} can be obtained from these transfer matrices
(in the homogeneous case) by means of trace identities. 
In this context, $B$ and $C$ are used respectively as  creation 
and annihilation operators of Bethe eigenstates.
The common eigenstates of the Hamiltonian and the commuting transfer matrices
are constructed by successive actions of operators $B(\la_j)$ on a reference
state $\ket{0}$, which in this case is the ferromagnetic state with all spins
up,  for any set of $n$ spectral parameters
$\{\lambda_j, 1 \le j \le n\}$ solution of Bethe equations.

\bigskip

The quantum inverse problem consists in reconstructing the local 
operators, which here are the local spin operators $S^\pm=S^x\pm iS^y$, $S^z$ 
at a given site $i$ of the chain, with $S^{\alpha} = {1 \over 2} \sigma^{\alpha}$, $\alpha = x, y, z$, $\sigma^{\alpha}$ being the standard Pauli matrices, in terms of the matrix elements $A$, $B$, 
$C$, $D$ of the monodromy matrix used to construct the eigenstates of the
Hamiltonian. This was done in \cite{KitMT99}, where we obtained the
following result :
\begin{align} 
 &S^-_i=\bigg\{\prod_{\alpha=1}^{i-1} (A+D)(\xi_{\alpha})\bigg\} 
               \, B(\xi_i) \,
     \bigg\{\prod_{\alpha=1}^{i} (A+D)(\xi_{\alpha})\bigg\}^{-1},
               \label{rec-Sm}\\
 &S^+_i =\bigg\{\prod_{\alpha=1}^{i-1} (A+D)(\xi_{\alpha})\bigg\} 
               \, C(\xi_i)  \,
     \bigg\{ \prod_{\alpha=1}^{i} (A+D)(\xi_{\alpha})\bigg\}^{-1},
               \label{rec-Sp}\\
 &S^z_i =\bigg\{\prod_{\alpha=1}^{i-1} (A+D)(\xi_{\alpha})\bigg\} 
               \, \frac{1}{2}  (A-D)(\xi_i)  \,
    \bigg\{ \prod_{\alpha=1}^{i} (A+D)(\xi_{\alpha})\bigg\}^{-1}.  
                \label{rec-Sz}
\end{align}
In other terms, the elementary operators $E_n^{\eps'_n,\eps_n}$, 
$\eps'_n,\eps_n \in \{1,2\}$, acting on the local quantum space
$\cH_n\simeq\Cset^2$ at site $n$ as the $2\times 2$ matrix 
$E_{ij}^{\eps',\eps}=\de_{i,\eps'}\de_{j,\eps}$, $1\le i,j \le 2$,
are reconstructed in the following way in terms of the matrix elements
$T_{\eps_n,\eps'_n}$ of the quantum monodromy matrix \eqref{mat-monodromie}:
\begin{equation}\label{reconst-1/2}
   E_n^{\eps'_n,\eps_n}=\bigg\{ 
      \prod_{\alpha=1}^{n-1} (A+D)(\xi_{\alpha})\bigg\} 
               \, T_{\eps_n,\eps'_n}(\xi_k)\,
      \bigg\{\prod_{\alpha=1}^{n} (A+D)(\xi_{\alpha})\bigg\}^{-1}.
\end{equation}

Note that the expressions \eqref{rec-Sm} - \eqref{reconst-1/2}, which are given here in the more general case with inhomogeneity parameters $\xi_j$, are in particular valid in the homogeneous limit where all parameters $\xi_j$ are set equals to zero.

These formulae were used in \cite{KitMT99} to obtain an explicit determinant
representation for the form factors of the finite chain, 
and in \cite{KitMT99a} to derive integral representations for general
correlation functions of the chain in a magnetic field at the thermodynamic
limit.

The proof of \eqref{rec-Sm}--\eqref{rec-Sz} proposed in \cite{KitMT99}
was very simple. It relies essentially on the fact that these formulae
can be easily proven, using that $R(0)=P$, for an operator
acting on the first site on the chain (cf. Lemma 4.2 of \cite{KitMT99}),
and that the problem can be reduced to this case by means of a propagator,
which is a shift operator from site 1 to site $i$, and which is given
as some product of $R$-matrices or as a product of transfer matrices
(cf. equations (4.3) and (4.4) of \cite{KitMT99}).  In \cite{KitMT99},
the propagator was also expressed in terms of factorizing $F$-matrices, but it is clear
that its properties rely actually only on its expression (equation (4.3) of
\cite{KitMT99}) as a product of $R$-matrices. Thus, the proof proposed in
\cite{KitMT99} is quite general, and can be extended to a large class of
lattice integrable models, as it will be shown in the next sections.
In particular, its generalisation to fundamental models is quite obvious (see section 4 for some other interesting examples),
but we will also show that one can obtain an explicit reconstruction for fused lattice models as well, such as for example the Heisenberg chains of higher spins (see section 5).



\section{General solution for lattice models}

In this section, we show that the resolution of the quantum inverse problem
given in \cite{KitMT99} in the case of the XXZ spin-1/2 Heisenberg chain
is quite general and can be extended to a large class of lattice models.

\bigskip

Let $\mathcal{A}$ be a Hopf algebra, for example a Yangian or a quantum affine algebra,
which admits a family of finite dimensional evaluation representations 
$W_r(\la)$, where $r+1$ is the dimension, depending on a spectral parameter
$\la$.
Suppose that, for any irreducible tensor product of two such representations
$W_r(\la)\otimes W_s(\mu)$, there exist a corresponding $R$-matrix
$R^{W_r W_s}(\la-\mu)$ solving the Yang-Baxter equation, and which is unitary (except for a finite number of values 
of the spectral parameter). 
Suppose moreover that the matrix $R^{W_r W_r}(\la)$, acting on the
tensor product of any two spaces with same dimension $r+1$, 
reduces to a permutation operator for
a particular value of the spectral parameter :
\begin{equation}\label{R0}
     R^{W_r W_r}(0)=P^{W_r W_r}.
\end{equation}
For notation conveniences, we will consider the additive spectral parameter case. However, it will be clear all along the results and proofs, that our method being purely algebraic, it extend to more general situations as well. 
Let us also suppose that there exists some fusion rules which allow the explicit construction, from the $R$-matrices $R^{V V}(\la)$ associated to fundamental representations $V(\la)$, of the $R$-matrices associated to higher dimensional
representations:
\begin{equation}\label{regle-fusion}
    R^{W_r W_s}(\la)=\cF^{W_r W_s, V}\left[R^{V V}(\la)\right].
\end{equation} 

Then let us consider a quantum lattice model, solvable by means of algebraic
Bethe ansatz, whose space of states is isomorphic to a tensor product
of evaluation representations:
\begin{equation*}
     \cH \simeq W_{r_1}(\xi_1)\otimes\dots\otimes W_{r_N}(\xi_n).
\end{equation*}
Here, $W_{r_n}(\xi_n)$ is the quantum local space at site $n$, and $\xi_n$ is
an inhomogeneity parameter attached to the site $n$. Let us define the quantum L-operator of the model at site $n$ by
\begin{align*}
    L_n(\la-\xi_n+\nu_0) &\equiv L^{VW_{r_n}}_{0n}(\la-\xi_n+\nu_0)\\
                  &= R^{V W_{r_n}}_{0n}(\la-\xi_n)
      \ \in \End(V(\la)\otimes W_{r_n}(\xi_n)),
\end{align*}
where $\nu_0\in\Cset$, 
the auxiliary space $V(\la)$ (labelled by subscript 0) being a fondamental
(evaluation) finite dimensional representation of $\mathcal{A}$.
The monodromy matrix, whose matrix elements are used to create eigenstates
of the Hamiltonian in the framework of algebraic Bethe ansatz, is given by
the ordered product of these $L$-operators along the chain:
\begin{equation}\label{mon-fond}
     T_{0,1\dots N}(\la)=L_{0N}(\la-\xi_N+\nu_0)\dots L_{01}(\la-\xi_1+\nu_0).
\end{equation}

In this context, the quantum inverse problem consists in reconstructing
local operators $E_n^\alpha$, $1\le \alpha\le (r_n+1)^2$, which generate
the local quantum space $\End(W_{r_n})$ at any given site $n$ of the chain,
in terms of the operator entries of the monodromy matrix \eqref{mon-fond}.

\medskip

The idea of this reconstruction, for an operator $E_n^\alpha$ acting at a
given site $n$ of the chain, {\em is to consider an auxiliary quantum $\cL$ operator for which the dimension of the auxiliary space
$W$ is equal to the dimension of the local quantum space $W_{r_n}$ at site $n$},
and such that this Lax operator at site $n$ reduces to the permutation operator
\eqref{R0} in a particular value of the spectral parameter. Thus, we define the following auxiliary Lax operator for any site $i$ of the chain, with an auxiliary matrix space $W$ isomorphic to the space of states $W_{r_n}$, for a given $n$, 
\begin{equation*}
   \cL_{0i}^{(r_n)} (\la-\xi_i+\nu_0) = R^{W W_{r_i}}_{0i}(\la-\xi_i),
\end{equation*}
acting in the tensor product $W\otimes W_{r_i}\simeq W_{r_n}\otimes W_{r_i}$, and such that it can be obtained by fusion from our starting quantum Lax operator $L_n$. 
Here and in the following, the upperscript $(r_n)$ means that the auxiliary space choosen to construct the corresponding object is isomorphic to $W_{r_n}$. 
For this special choice of the auxiliary space, let us note 
$T^{(r_n)} (\la)$ and $t^{(r_n)}(\la)$ the corresponding monodromy and
transfer matrices:
\begin{align}\label{mon-grand}
   &T^{(r_n)} (\la)\equiv T_{0,1\dots N}^{(r_n)} (\la)
       =\cL_{0N}^{(r_n)} (\la-\xi_N+\nu_0)\dots\cL_{01}^{(r_n)} (\la-\xi_1+\nu_0),\\
   &t^{(r_n)}(\la)\equiv t_{1\dots N}^{(r_n)}(\la)
                        =\tr_0\left(T_{0,1\dots N}^{(r_n)} (\la)\right).
\end{align}
We have the following proposition:

\begin{prop}\label{prop-reconstr}
  An operator $E_n^\alpha\in\End(W_{r_n})$ in a given site $n$ of the chain
  can be expressed in the following way in terms of the monodromy matrix
  $T_{0,1\dots N}^{(r_n)} (\xi_n)$ constructed with an auxiliary space $W\simeq W_{r_n}$ and of the transfer matrices constructed with the different auxiliary spaces isomorphic to the different quantum local space of states $W_{r_i}$ for any site index $i$, 
  $t^{(r_i)}(\xi_i)$, $1\le i \le n$:
  \begin{equation}\label{rec}
     E_n^\alpha= \bigg\{ \prod_{i=1}^{n-1} t^{(r_i)}(\xi_i)\bigg\}\,
           \tr_0 \left( E_0^\alpha\, T_{0,1\dots N}^{(r_n)} (\xi_n)\right)\,
             \bigg\{ \prod_{i=1}^{n} t^{(r_i)}(\xi_i)\bigg\}^{-1},
  \end{equation}
  where $E_0^\alpha$ denotes the operator $E_n^\alpha$ acting in the
  auxiliary space $W\simeq W_{r_n}$.
\end{prop}

\proof
It comes directly from the two following lemmas.

\begin{lemme}
The product of the above defined transfer matrices (which all commute among themselves due to the Yang-Baxter relation), 
\begin{equation*}
     T_{1\rightarrow n}=\prod_{i=1}^{n-1} t^{(r_i)}(\xi_i)
\end{equation*}
is an invertible shift operator from site 1 to site $n$, whose
action on the monodromy matrix (for any choice of the auxiliary space labelled by $0'$)
is given by
\begin{equation}\label{trans}
     T_{1\rightarrow n}\, T_{0',1\dots N}(\la)\, T_{1\rightarrow n}^{-1}
                       = T_{0',n\dots N1\dots n-1}(\la).
\end{equation}
\end{lemme}

\proof
The expression \eqref{trans} can be proved by induction on $n$.
The transfer matrix $t^{(r_n)}_{1\dots N}(\xi_n)$ constructed with auxiliary space $W$ isomorphic to $W_{r_n}$ can be expressed as a 
product of R-matrices:
\begin{align*}
   t^{(r_n)}_{1\dots N}(\xi_n)
  &=tr_0\Bigl( R_{0N}^{W W_{r_N}}(\xi_n-\xi_N)\dots 
         R_{0\, n+1}^{W W_{r_{n+1}}}(\xi_n-\xi_{n+1})\,P^{W W_{r_n}}_{0 n}\\
  &\qquad\qquad\qquad\times
         R_{0\, n-1}^{W W_{r_{n-1}}}(\xi_n-\xi_{n-1})\dots  
         R_{01}^{W W_{r_1}}(\xi_n-\xi_{1})\Bigr)\\
  &=tr_0\Bigl(P^{W W_{r_n}}_{0 n}\, 
         R_{n\, n-1}^{W_{r_n} W_{r_{n-1}}}(\xi_n-\xi_{n-1})
         \dots R_{n 1}^{W_{r_n} W_{r_1}}(\xi_n-\xi_{1})\\
  &\qquad\qquad\qquad\times  
         R_{nN}^{W_{r_n} W_{r_N}}(\xi_n-\xi_N)\dots 
         R_{n\, n+1}^{W_{r_n} W_{r_{n+1}}}(\xi_n-\xi_{n+1})\Bigr)\\
  &=R_{n\, n-1}^{W_{r_n} W_{r_{n-1}}}(\xi_n-\xi_{n-1})
         \dots R_{n 1}^{W_{r_n} W_{r_1}}(\xi_n-\xi_{1})\\
  &\qquad\qquad\qquad\times 
         R_{nN}^{W_{r_n} W_{r_N}}(\xi_n-\xi_N)\dots 
         R_{n\, n+1}^{W_{r_n} W_{r_{n+1}}}(\xi_n-\xi_{n+1}),
\end{align*}
where we used the cyclicity of the trace and the fact that, as $W\simeq W_{r_n}$,
$R_{0n}^{W W_{r_n}}(0)$ is the permutation $P^{W W_{r_n}}_{0 n}$ of the
auxiliary space $W$ and the quantum local space at site $n$, $W_{r_n}$.
Using also the expression of the monodromy matrix, for an auxiliary space
$W'$ of arbitrary dimension, as a product of $R$-matrices,  
\begin{multline*}
   \esp T_{0',n\dots N1\dots n-1}(\la)
        =R_{0'n-1}^{W' W_{r_{n-1}}}(\la-\xi_{n-1})\dots R_{0'1}^{W' W_{r_1}}(\la-\xi_{1})\\
         \times R_{0'N}^{W' W_{r_N}}(\la-\xi_N)\dots 
            R_{0'\, n}^{W' W_{r_{n}}}(\la-\xi_{n}),\qquad
\end{multline*}
one can easily show, by using successively  the Yang-Baxter equation
\begin{equation*}
  R_{ni} (\xi_n - \xi_i)\ R_{0'i} (\lambda - \xi_i)\ 
      R_{0'n} (\lambda -\xi_n) = 
  R_{0'n} (\lambda -\xi_n)\ R_{0'i} (\lambda - \xi_i)\ 
      R_{ni} (\xi_n - \xi_i),
\end{equation*}
for $i = n+1, \dots, N, 1, \dots, n-1$ that,
\begin{equation*}
   t^{(r_n)}_{1\dots N}(\xi_n)\, T_{0',n\dots N1\dots n-1}(\la) 
     =T_{0',n+1\dots N1\dots n}(\la)\, t^{(r_n)}_{1\dots N}(\xi_n),
\end{equation*}
for any $n$, which achieves the proof.
\cqfd

\begin{lemme}
   If the auxiliary space  $W$ has same dimension as the quantum space
   $W_{r_1}$ at site 1, one has the following identity, for any operator
   $E^\alpha\in\End(W_{r_1}\simeq W)$:
   \begin{equation*}
      \tr_0 \left( E_0^\alpha\, T_{0,1\dots N}^{(r_1)} (\xi_1)\right)
       = E_1^\alpha\, t^{(r_1)}_{1\dots N}(\xi_1).
   \end{equation*}
\end{lemme}

\proof
Starting from the expression of $T_{0,1\dots N}^{(r_1)} (\xi_1)$ as a product of
$R$-matrices, on uses the fact that $R_{01}^{W W_{r_1}}(0)$ is the permutation
operator $P_{0 1}$, which acts in the following way on $E_0^\alpha$: 
$P_{0 1}\, E_0^\alpha=E_1^\alpha\,P_{0 1} $. 
\cqfd

It should be noted that we have considered here the general inhomogeneity case, but of course, our result applies and simplifies for the homogeneous case where all parameters $\xi_n$ are equal.

\bigskip

Proposition \ref{prop-reconstr} gives thus the reconstruction of local 
operators at site $n$ in terms of the matrix elements of the monodromy matrix
{\em the auxiliary space of which has the same dimension as the quantum
space of states at site $n$}.
But the creation and annihilation operators used when solving the model by
means of algebraic Bethe ansatz are elements of the monodromy matrix
 $T_{1\dots N}(\la)$ \eqref{mon-fond} {\em the auxiliary space of which is a 
fundamental representation} of the algebra $\mathcal{A}$. 
One can use then the {\em fusion rules} \eqref{regle-fusion} to reexpress
the matrix elements of the monodromy matrix \eqref{mon-grand}, which appear in
the expression \eqref{rec} of proposition \ref{prop-reconstr} in terms of
those of \eqref{mon-fond}, solving the problem. We will see some example in the next sections.
 


\section{Fundamental lattice models}

In this section, we apply the general formula of section 3 in the special case of fundamental lattice models. Fundamental lattice models are characterized by the fact that the quantum space of states at any site $n$ of the lattice is given by the fundamental representation of the corresponding quantum algebra and is isomorphic to the auxiliary matrix space. Hence, the Lax operator is  just given by the quantum $R$-matrix in the tensor product of the fundamental representation by itself and indeed reduces to the permutation operator in this tensor product when the spectral parameter is equal to zero,
\begin{align*}
    L_n(\la, \xi_n) &\equiv L^{V V_n}_{0n}(\la-\xi_n)\\
                  &= R^{V V}_{0n}(\la-\xi_n)
      \ \in \End(V(\la)\otimes V(\xi_n)),
\end{align*}
where $V$ is a fundamental representation. The space of states is here
\begin{equation*}
     \cH \simeq V(\xi_1)\otimes\dots\otimes V(\xi_N).
\end{equation*}
The quantum monodromy matrix is given by
\begin{equation*}
     T_{0, 1\dots N}(\la)=L_{0n}(\la, \xi_n)\dots L_{01}(\la, \xi_1),
\end{equation*}
and the transfer matrix is defined as usual by its trace in auxiliary space $V_0$, 
\begin{equation*}
t_{1\dots N}(\la) = \tr_0\left(T_{0,1\dots N} (\la)\right).
\end{equation*}
As a consequence, the formula \eqref{rec} of Proposition \ref{prop-reconstr} applies directly to give the reconstruction of any local operator $E_n^\alpha$ acting at a
given site $n$ of the chain,
 \begin{equation}\label{rec-fond}
     E_n^\alpha= \bigg\{ \prod_{i=1}^{n-1} t (\xi_i)\bigg\}\,
           \tr_0 \left( E_0^\alpha\, T_{0,1\dots N} (\xi_n)\right)\,
             \bigg\{ \prod_{i=1}^{n} t(\xi_i)\bigg\}^{-1},
  \end{equation}
  where $E_0^\alpha$ denotes the operator $E_n^\alpha$ acting in the
  auxiliary space. Note that the homogeneous case is obtained as usual by putting all inhomogeneity parameters $\xi_n$ to zero in the above formula.

Let us stress here that it is a very direct copy of our original formula 
\eqref{reconst-1/2} given in \cite{KitMT99} for the case of the $XXZ$ spin-$1 \over 2$  chain, which is completely generic in this respect. This formula is of course valid for both inhomogemeous and homogeneous models which are obtained simply in the limit where all inhomogeneity parameters $\xi_n$ are set equal to zero.

\subsection{The $sl_{n+1}$ XXZ model}

The quantum $R$-matrix of this model, solving the Yang-Baxter equation and associated to a fundamental representation of $\mathcal{U}_q(sl_{n+1})$, is defined by the following $(n+1)^2\times (n+1)^2$ matrix \cite{Che80,BabVV81,PerS81},
\begin{equation}\label{mat-Rn}
    R^{(n)}(\la)=\frac{e^\la\mathrm{R}^{(n)}-e^{-\la}\bar{\mathrm{R}}^{(n)}}
                {e^\la q-e^{-\la}q^{-1}},\qquad (q=e^\eta)
\end{equation}
with,
\begin{align*}
   &\mathrm{R}^{(n)}
     =\sum_{\alpha\ne \beta} E^{\alpha\alpha}\otimes E^{\beta\beta}
     +q\sum_\alpha E^{\alpha\alpha}\otimes E^{\alpha\alpha}
     +(q-q^{-1}) \sum_{\alpha<\beta} E^{\alpha\beta}\otimes E^{\beta\alpha},\\
   &\bar{\mathrm{R}}^{(n)}
     =\sum_{\alpha\ne \beta} E^{\alpha\alpha}\otimes E^{\beta\beta}
     +q^{-1}\sum_\alpha E^{\alpha\alpha}\otimes E^{\alpha\alpha}
     -(q-q^{-1}) \sum_{\alpha<\beta} E^{\alpha\beta}\otimes E^{\beta\alpha},
\end{align*}
where $E^{\alpha\beta}$ is the elementary $(n+1)\times (n+1)$-matrix with only one non-zero element, $E^{\alpha\beta}_{ij}=\de_{i,\alpha}\de_{j,\beta}$. Note that, $R(0)=P$, the permutation operator. For homogeneous models, the quantum Lax operator being directly equal to this $R$-matrix, the quantum monodromy matrix is given by,
\begin{equation}\label{Tn}
   T_{0,1\dots N}^{(n)}(\la)= R_{0N}^{(n)}(\la)\dots 
                        R_{02}^{(n)}(\la) R_{01}^{(n)}(\la).
\end{equation}
The transfer matrices $t^{(n)}(\la)=\tr_0  T_{0,1\dots N}^{(n)}(\la)$ commute among themselves for arbitrary values of the spectral parameters $\la$, and one can define the local quantum Hamiltonian by means of trace identities,
\begin{equation}\label{Ham-n}
  \H^{(n)}= (q-q^{-1})\frac{d}{d\la}\ln t^{(n)}(\la)\pour{\la=0}-N(q+q^{-1}).
\end{equation}
The simultaneous diagonalization of the transfer matrix and of the Hamiltonian is given by the so-called nested Bethe ansatz \cite{Sut75,KulR81a}. 
The solution of the quantum inverse scattering problem consists in reconstructing the local operators 
\begin{equation*}
   E_p^{\eps'_p,\eps_p}, \qquad 1\le \eps'_p,\eps_p \le n+1,
\end{equation*}
acting in the local quantum space at site $p$, $\cH_p\simeq \Cset^{n+1}$, as the elementary matrices $E^{\eps'_p,\eps_p}_{ij}=\de_{i,\eps'_p}\de_{j,\eps_p}$, in terms of the elements of the quantum monodromy matrix $T^{(n)}_{0,1\dots N}(\la)$, whose elements $T_{1j}(\la)$, $2\le j\le n+1$ are used in particular to create Bethe eigenstates. We have directly from the above formula,
\begin{equation}
     E_p^{\eps'_p,\eps_p}= 
           \bigg\{t^{(n)}(0)\bigg\}^{p-1}\,
       \tr_0 \left( E_0^{\eps'_p,\eps_p}\, T_{0,1\dots N}^{(n)} (0)\right)\,
             \bigg\{ t^{(n)}(0)\bigg\}^{-p}.
\end{equation}
where the trace can be computed in terms of the matrix elements $(T_{0,1\dots N}^{(n)} (0))_{\eps_p,\eps'_p}$ of the quantum monodromy matrix \eqref{Tn} to give finally,
\begin{equation}\label{reconst-n}
     E_p^{\eps'_p,\eps_p}= 
           \bigg\{t^{(n)}(0)\bigg\}^{p-1}\,
        \left( T_{0,1\dots N}^{(n)} (0)\right)_{\eps_p,\eps'_p}\,
             \bigg\{ t^{(n)}(0)\bigg\}^{-p}.
\end{equation}
This expression \eqref{reconst-n} gives the reconstruction of the quantum local operators for the homogeneous chain and is a direct generalization of \eqref{reconst-1/2}. It is clear from the general formulas of the preceding section that it is also possible to write a similar formula for the inhomogeneous case as well.

\subsection{The XYZ spin-$1 \over 2$ model}
The XYZ model for spin-$1 \over 2$ defined on a chain of lenght $N$ is given by the following Hamiltonian operator,
\begin{equation}
H_{XYZ} = \sum_{m=1}^N \big\{ J_x \sigma_m^x \sigma^x_{m+1} +
 J_y \sigma^y_m\sigma^y_{m+1} + J_z \sigma^z_m\sigma^z_{m+1} \big\}.
\label{eq:Hxyz} 
\end{equation}
This model is associated to a quantum $R$-matrix of elliptic type, given by \cite{Bax72},
\begin{equation}\label{Rxyz}
   R(\lambda, \mu)=
      \begin{pmatrix}
         1 & 0 & 0 & d(\lambda -\mu)\\
         0 & b(\lambda - \mu) & c(\lambda - \mu) & 0\\
         0 & c(\lambda - \mu) & b(\lambda - \mu) & 0\\
         d(\lambda - \mu) & 0 & 0 & 1
      \end{pmatrix}
\end{equation}
with,
\begin{align}
  b(u) & = \frac{\sn(u)}{\sn(u+\eta)},\\
  c(u) & = \frac{\sn(\eta)}{\sn(u+\eta)},\\
  d(u) & = k \sn(\eta) \sn(u),
\end{align}
where $\sn(u) \equiv \sn(u;k)$ is the jacobi elliptic function of modulus $k$, $0 \le k \le 1$. The constants in the Hamiltonian are related to the parameters as,
\begin{align}
  J_x & = 1 + k \sn^2 (\eta),\\
  J_y & = 1 - k \sn^2 (\eta),\\
  J_z & = \cn(\eta) \dn(\eta).
\end{align}

Then it is obvious that this $R$-matrix reduces to the permutation operator $P$ for $u=0$. Moreover it is unitary up to a numerical factor that can be taken into account in all our computations. Defining the quantum Lax operator as being the $R$-matrix itself, all the requirements for applying our above general formula are satisfied. Hence, we just obtain the reconstruction of any spin operator at site $n$ to be given {\em by the same formula as in the XXZ case} where we have just to replace the operators $A, B, C, D$ in \eqref{reconst-1/2} by the corresponding one's from the $XYZ$ model.



\section{Quantum inverse problem for higher spin XXX chains}

In this section, we consider the integrable generalization of the XXX model with local 
spin variables $S_n^\alpha$ realizing a 
representation of $sl_2$ of dimension $l+1$ (spin $l/2$).
This model is given by  a Lax operator $L^{(1)}_{0n}(\la)$, 
acting on the tensor product of a 2-dimensional auxiliary space $V_0$
by the local quantum space $\cH_n\simeq\Cset^{l+1}$ at site $n$,
\begin{equation}\label{op-L-l}
   L_{0n}^{(1)}(\lambda)=\frac{1}{\la+\eta l}
         \begin{pmatrix}
             \lambda+\eta(\frac{l}{2}+ S_n^z) & \eta S_n^-\\
             \eta S_n^+ & \lambda+\eta(\frac{l}{2}- S_n^z)
         \end{pmatrix}.
\end{equation}
As in the spin 1/2 case, the corresponding monodromy matrix, given by the 
ordered product of these
$L$-operators with 2-dimensional auxiliary space along the chain,
\begin{equation}\label{mat-mon1}
   T^{(1)}_{0,1\dots N} (\lambda) = 
          L_{0N}^{(1)}(\lambda)\dots L_{01}^{(1)}(\lambda),
\end{equation}
can be written on the auxiliary space $V_0$ as a $2\times 2$ matrix, 
\begin{equation}\label{mat-mon}
   T^{(1)}_{0,1\dots N} (\lambda) = 
      \begin{pmatrix}
         A (\lambda) & B (\lambda)\\
         C (\lambda)& D (\lambda)
      \end{pmatrix}, 
\end{equation}
whose matrix elements are used to construct 
eigenstates of the Hamiltonian.
The commutation relations \eqref{commutation} of these matrix elements are given by the same
$R$-matrix \eqref{mat-R} as in the spin 1/2 case. 

The Hamiltonian of this model is given by means of trace identities involving the
transfer matrices $t^{(l)}(\la)$ constructed from $L$-operators 
$L_{0n}^{(l)}(\la)$ with $(l+1)$-dimensional auxiliary space.
This Hamiltonian describes nearest-neighbour interactions, which follows from
the fact that  $L^{(l)}_{0n}(0)$ reduces to the permutation $P_{0n}$.
The $L$-operators $L_{0n}^{(s)}(\la)$ with higher dimensional $(s+1)$-auxiliary
space can be constructed from \eqref{op-L-l} by the fusion procedure
\cite{KulRS81}. More precisely, if $P^+_{a_1\dots a_s}$ is the projector 
(symmetrizer) onto
the $(s+1)$-dimensional subspace $V_{(a_1\dots a_s)}$ of the tensor product
$V_{a_1}\otimes \dots\otimes V_{a_s}$ of $s$ auxiliary spaces 
$V_{a_i}\simeq \Cset^2$, one has
\begin{multline}\label{fus-L}
   \esp L^{(s)}_{(a_1\dots a_s),n} (\la)=
    P^+_{a_1\dots a_s} \,
             L^{(1)}_{a_1,n} \Bigl(\la+\frac{s-1}{2}\eta\Bigr)\dots\\
              \dots   L^{(1)}_{a_j,n} \Bigl(\la-\frac{s+1-2j}{2}\eta\Bigr)\dots
                 L^{(1)}_{a_s,n} \Bigl(\la-\frac{s-1}{2}\eta\Bigr) \,
    P^+_{a_1\dots a_s}.
\end{multline}
Remember that, in all this section, the dimension of the local quantum spaces is fixed to be
$l+1$, and that the upperscript $(s)$ labels only the dimension of the auxiliary
space (to be $s+1$).

\bigskip

The quantum inverse problem consists here in reconstructing the local spin
operators $S_n^\alpha$ in terms of the matrix elements $A$, $B$, $C$, $D$
of the monodromy matrix \eqref{mat-mon} with 2-dimensional auxiliary space. 
From Proposition \ref{prop-reconstr}, we know how to reconstruct them
in terms of the matrix elements of the monodromy matrix $T^{(l)}(\la)$
with auxiliary space of dimension $l+1$:
  \begin{equation}\label{recl}
     S_n^\alpha= \left\{ t^{(l)}(0)\right\}^{n-1}\,
           \tr_0 \left( S_0^\alpha\, T_{0,1\dots N}^{(l)} (0)\right)\,
             \left\{ t^{(l)}(0)\right\}^{-n}.
  \end{equation}
The problem is thus to express the quantities
\begin{equation}\label{lambda}
   \La_\alpha^{(s)}(\la)=
         \tr_0 \left( S_0^\alpha\, T_{0,1\dots N}^{(s)} (\la)\right)
\end{equation}
for any component $\alpha$ of the spin operator and for any integer $s\le l$.
To do this, one uses the fact that the monodromy matrix  $T_{0,1\dots N}^{(s)}$
with $(s+1)$-auxiliary space can be obtained by 
fusion in terms of the monodromy matrix $T_{0,1\dots N}^{(1)}$
with 2-dimensional auxiliary space, similarly as in \eqref{fus-L}:
\begin{multline}\label{fus-T}
   \esp T^{(s)}_{(a_1\dots a_s),1\dots N} (\la)=
    P^+_{a_1\dots a_s}\, 
           T^{(1)}_{a_1,1\dots N} \Bigl(\la+\frac{s-1}{2}\eta\Bigr)\dots\\
     \dots T^{(1)}_{a_j,1\dots N} \Bigl(\la-\frac{s+1-2j}{2}\eta\Bigr)\dots
           T^{(1)}_{a_s,1\dots N} \Bigl(\la-\frac{s-1}{2}\eta\Bigr)\,
    P^+_{a_1\dots a_s}.
\end{multline}
This fusion procedure enables us to formulate the following lemma:

\begin{lemme}
   For any integer $s\le l$, the quantity $\La_\alpha^{(s)}(\la)$ defined by
equation \eqref{lambda} has the following expression in terms of the elements
$\La_\alpha^{(1)}(\la)$ of the monodromy matrix \eqref{mat-mon} and of the
transfer matrices $t^{(s')}(\la)$ corresponding to $(s'+1)$-dimensional
auxiliary space ($s'\le s$):
\begin{equation}\label{form}
  \La_\alpha^{(s)}(u)=\sum_{k=1}^s t^{(s-k)}\Bigl(u+\frac{k}{2}\eta\Bigr)\,
                      \La_\alpha^{(1)}\Bigl(u+\frac{2k-s-1}{2}\eta\Bigr)\,
                          t^{(k-1)}\Bigl(u+\frac{k-s-1}{2}\eta\Bigr).
\end{equation}
\end{lemme}

\proof
We will show the formula \eqref{form} by recursion on $s$, for $s\le l$.
First, note that it is satisfied for $s=1$. Moreover, it can be easily
shown also for $s=2$.

Suppose now that it is satisfied until a given $s<l$. 
When fusioning a spin $s/2$ with spin $1/2$ representation, one has the
following block-triangular decomposition for the product of the corresponding monodromy matrices
\cite{KulR82,KirR87}:
\begin{multline}\label{bloc}
  \esp T_{a_1,1\dots N}^{(1)}\bigl(u+\frac{s+1}{2}\eta\bigr)\,
   T_{a_2,1\dots N}^{(s)}(u)= \\
     = \begin{pmatrix}
         T_{(a_1 a_2),1\dots N}^{(s+1)}(u+\frac{\eta}{2}) & 0 \\
         * & f_l(u+\frac{s}{2}\eta)
            T_{\langle a_1 a_2 \rangle ,1\dots N}^{(s-1)}(u-\frac{\eta}{2})
      \end{pmatrix},
\end{multline} 
where the block structure is in accordance with the decomposition
\begin{alignat*}{3}
    &V_{a_1}\otimes V_{a_2}\simeq  V_{(a_1a_2)}\oplus V_{\langle a_1 a_2 \rangle},
     & \qquad  &V_{a_1}\simeq \Cset^2, &\quad  &V_{a_2}\simeq \Cset^{s+1},\\
     & & \qquad  &V_{(a_1a_2)}\simeq \Cset^{s+2}, &\quad
          &V_{\langle a_1 a_2 \rangle} \simeq \Cset^{s},    
\end{alignat*}
and where the function $f_l(u)$ is the quantum determinant of the monodromy
matrix \eqref{mat-mon}.
Taking the trace in \eqref{bloc}, one obtains the well-known fusion equation
for the transfer matrices:
\begin{equation}\label{rec-trans}
   t^{(1)} (u+\frac{s+1}{2}\eta)\,  t^{(s)} (u)=
   t^{(s+1)} (u+\frac{\eta}{2})+ 
   f_l(u+\frac{s}{2}\eta)\, t^{(s-1)}(u-\frac{\eta}{2}).
\end{equation} 
In the same way, decomposing the component of the spin
$S^\alpha_{a_1 a_2}=S^\alpha_{a_1}+S^\alpha_{a_2}$ into 
$S^\alpha_{(a_1 a_2)}+S^\alpha_{\langle a_1 a_2\rangle}$ on 
$V_{(a_1a_2)}\oplus V_{\langle a_1 a_2 \rangle}$, one can take the trace
of the quantity 
\begin{equation*}
   S^\alpha_{a_1 a_2}\, T_{a_1,1\dots N}^{(1)}(u+\frac{s+1}{2}\eta)\,
   T_{a_2,1\dots N}^{(s)}(u)
\end{equation*}
in \eqref{bloc}, which leads to the following recursion relation for 
$\La_\alpha^{(s)}(u)$:
\begin{multline}\label{rec-la}
   \esp\La_\alpha^{(1)}(u+\frac{s+1}{2}\eta)\,t^{(s)}(u)+
             t^{(1)}(u+\frac{s+1}{2}\eta)\,\La_\alpha^{(s)}(u)\\
   =
   \La_\alpha^{(s+1)}(u+\frac{\eta}{2})+
   f_l(u+\frac{s}{2}\eta)\,\La_\alpha^{(s-1)}(u-\frac{\eta}{2}).
\end{multline}
One can then easily show, replacing $\La_\alpha^{(s)}(u)$ and 
$\La_\alpha^{(s-1)}(u)$ by the corresponding expressions
\eqref{form} in \eqref{rec-la}, and using the recursion relation 
\eqref{rec-trans} to reconstruct the transfer matrices which appear in 
each term of the sum, that one obtains also the expression \eqref{form} for 
$\La_\alpha^{(s+1)}(u)$.
\cqfd

Consequently, the solution of the quantum inverse problem
for local spin operators is given by the following proposition:

\begin{prop}
   The local spin operators $S_n^z$, $S^\pm_n=S^x_n\pm iS^y_n$ at site $n$
have the following expressions in terms of the matrix elements of the monodromy
matrix \eqref{mat-mon}:
\begin{align} 
 &S^-_i=\left\{ t^{(l)}(0)\right\}^{i-1} 
                              \nonumber\\
       &\qquad\times \bigg\{ 
             \sum_{k=1}^l t^{(l-k)}\Bigl(\frac{k}{2}\eta\Bigr)\,
                                B\Bigl(\frac{2k-l-1}{2}\eta\Bigr)\,
                               t^{(k-1)}\Bigl(\frac{k-l-1}{2}\eta\Bigr)\bigg\}
        \left\{ t^{(l)}(0)\right\}^{-i}, 
                           \label{spinlm}
                \displaybreak[0]\\
 &S^+_i =\left\{ t^{(l)}(0)\right\}^{i-1}
                              \nonumber\\
       &\qquad\times \bigg\{ 
              \sum_{k=1}^l t^{(l-k)}\Bigl(\frac{k}{2}\eta\Bigr)\,
                            C\Bigl(\frac{2k-l-1}{2}\eta\Bigr)\,
                            t^{(k-1)}\Bigl(\frac{k-l-1}{2}\eta\Bigr)\bigg\}
         \left\{t^{(l)}(0)\right\}^{-i}, 
                           \label{spinlp}\displaybreak[0]\\    
 &S^z_i =\left\{ t^{(l)}(0)\right\}^{i-1} 
                              \nonumber\\
       &\qquad\times \bigg\{ 
              \sum_{k=1}^l t^{(l-k)}\Bigl(\frac{k}{2}\eta\Bigr)\,
                        \frac{1}{2}(A-D)\Bigl(\frac{2k-l-1}{2}\eta\Bigr)\,
                            t^{(k-1)}\Bigl(\frac{k-l-1}{2}\eta\Bigr)\bigg\}
        \left\{ t^{(l)}(0)\right\}^{-i}.
                          \label{spinlz}
\end{align}
\end{prop}

Note that the above formulae can be easily generalized to an inhomogeneous 
model.

\begin{rem}
The transfer matrices $t^{(s)}(u)$ can be expressed (using \eqref{rec-trans})
in terms of $t^{(1)}(u)$ and the quantum determinant (see \cite{KulS82,KulR82,KirR87}).
\end{rem}

\begin{rem}
   The proof given here being quite general and relying essentially on the 
recursion relation \eqref{bloc} with respect to the dimension of the auxiliary 
space, similar results can be obtained for XXZ and XYZ higher spin chains.
\end{rem}

\bigskip

It is quite interesting to examine the quasi-classical limit of the above formulas. This amounts to consider the limit in which the quantum parameter $\eta$ goes to zero while the spin $s$ of the quantum representation space goes to infinity, their product (the classical spin) being kept fixed. In the above formulas, the sum on spin and shifts becomes an integral in this limit. 

In this context, it is worth mentioning that while the solution to the quantum inverse scattering problem admits a very simple solution (essentially as a product of elements of the quantum monodromy matrix), the semi-classical limit of this expression (which is equivalent to the Gel'fand-Levitan-Marchenko equations) is more involved. This is in particular due to the fact that while every term in the multiplicative reconstruction formula for the quantum model becomes in this limit a function of the classical spin, and commutes with any other term in this expression, there is an ``accumulation'' of such terms which number becomes of the order $s$ (i.e, of order $1 \over {\eta}$ in the limit ${\eta} \rightarrow 0$). Hence, this gives non-trivial contribution (coming from non-commutative effects) in the classical limit , explaining in some sense why the classical case seems a priori much more involved than the quantum one. These effects results in effective ``universal'' Backlund transformations coming from the multiple dressing of the $A, B, C, D$ operators in the quantum formulas by transfer matrices. It also shed some light on the reason why the direct quantization of the Gel'fand-Levitan-Marchenko equations is a priori a very difficult problem to deal with, since it would be very hard to guess the mutiplicative quantum formula from it. We expect do discuss this classical limit in a separate publication.



\section{Conclusion}

We have shown in this article that the quantum inverse scattering problem can be solved for almost all known quantum integrable lattice models with arbitrary finite dimensional local space of states (which could be different from one site to another). Our derivation relies on the unitaity of the quantum $R$-matrix, its initial condition, $R(0) = P$, and on the existence of fusion rules relating quantum Lax operators associated to different auxiliary spaces. Hence, our results apply not only to the fundamental lattice models (such as the Heisenberg spin chains and their higher rank algebra generalisations) but also to lattice models with arbitrary number of impurities and to the so-called fused lattice models, like the higher spin integrable generalisations of Heisenberg chains. 
We would like to stress that besides some elementary technical details, the fundamental lattice model situation, both in the result and in its proof, is a very direct copy of our original derivation for the $XXZ$ spin$1 \over 2$ case \cite{KitMT99}. For other more general models, fusion should be used, leading to nice dressed formulas such as \eqref{spinlm}--\eqref{spinlz}.
Another interesting case, that we have not considered here, concerns quantum integrable lattice field theory models, such as the Sine-Gordon model. In this context, we will have to deal with auxiliary quantum Lax operators having infinite dimensional representation auxiliary space, and also to define a trace operation for such representations. We expect the resulting reconstruction expressions for the local quantum fields to be given by formulas very similar to the infinite spin limit of the above result for spin-$s$ Heisenberg chains.

These results can now be applied to the computation of form factors and correlation functions of these models along the line we described in \cite{KitMT99a}.

As a last remark, we note that after the completion of this manuscript, a paper,  strongly inspired by our original work \cite{KitMT99}, and the seminar given by one of us \cite{Mai99}, appeared in the hep-th/preprint archive (hep-th/9910253), also dealing with the quantum inverse problem, however only in the particular case of fundamental graded  lattice models.  

\vskip 1cm
{\bf\large Acknowlegement.} We would like to thank N. Kitanine for several useful discussions and remarks. We also would like to thank O. Babelon, A. Izergin, N. Yu. Reshetikhin, E. K. Sklyanin, F. Smirnov and all our colleages of the Theory Group in ENS Lyon for discussions and for their interest in this work. We also thank the CRM in Montreal for hospitality during the ``S\'eminaire de Math\'ematique Sup\'erieures'' in July-August 1999, and where the results of section 5 were obtained.




\end{document}